\documentclass[twocolumn]{aastex61}

\usepackage{amsmath}

\newcommand{\cms}{\mbox{cm s$^{-1}$}}
\newcommand{\ms}{\mbox{m s$^{-1}$}}
\newcommand{\kms}{\mbox{km s$^{-1}$}}

\received{December 17, 2016}
\revised{July 26, 2017}
\accepted{July 26, 2017}
\submitjournal{\apj}

\begin{document}

\title{Insights on the spectral signatures of stellar activity and planets from PCA}

\shorttitle{Signatures of Stellar Activity and Planets from PCA}
\shortauthors{Davis et al.}

\author[0000-0002-5070-8395]{Allen B. Davis}
\affiliation{Department of Astronomy, Yale University, 52 Hillhouse Ave, New Haven, CT 06511, USA}

\author[0000-0002-9656-2272]{Jessi Cisewski}
\affiliation{Department of Statistics, Yale University, 24 Hillhouse Ave, New Haven, CT 06511, USA}

\author[0000-0002-9332-2011]{Xavier Dumusque}
\affiliation{Observatoire de Gen\`eve, Universit\'e de Gen\`eve, 51 ch. des Maillettes, 1290 Versoix, Switzerland}
\affiliation{Harvard-Smithsonian Center for Astrophysics, 60 Garden Street, Cambridge, MA 02138, USA}

\author[0000-0003-2221-0861]{Debra A. Fischer}
\affiliation{Department of Astronomy, Yale University, 52 Hillhouse Ave, New Haven, CT 06511, USA}

\author[0000-0001-6545-639X]{Eric B. Ford}
\affiliation{Center for Exoplanets and Habitable Worlds, The Pennsylvania State University, University Park, PA 16802, USA}
\affiliation{Center for Astrostatistics, The Pennsylvania State University, University Park, PA 16802, USA}
\affiliation{Institute for CyberScience, The Pennsylvania State University, University Park, PA 16802, USA}
\affiliation{Department of Astronomy \& Astrophysics, The Pennsylvania State University, University Park, PA 16802, USA}

\correspondingauthor{Allen B. Davis}
\email{allen.b.davis@yale.edu}


\begin{abstract}

Photospheric velocities and stellar activity features such as spots and faculae produce measurable radial velocity signals that currently obscure the detection of sub-meter-per-second planetary signals. However, photospheric velocities are imprinted differently in a high-resolution spectrum than Keplerian Doppler shifts. Photospheric activity produces subtle differences in the shapes of absorption lines due to differences in how temperature or pressure affects the atomic transitions. In contrast, Keplerian Doppler shifts affect every spectral line in the same way. With high enough S/N and high enough resolution, statistical techniques can exploit differences in spectra to disentangle the photospheric velocities and detect lower-amplitude exoplanet signals. We use simulated disk-integrated time-series spectra and principal component analysis (PCA) to show that photospheric signals introduce spectral line variability that is distinct from Doppler shifts. We quantify the impact of instrumental resolution and S/N for this work.

\end{abstract}

\section{Introduction}

The search for exoplanets is one of the most exciting scientific pursuits of this century. In the past 20 years, hundreds of exoplanets have been detected using the Doppler (or radial velocity; RV) technique. These discoveries have inspired booming new subfields in astronomy: exoplanet detection and characterization. NASA's Kepler Mission \citep{Borucki2010} stopped just short of deriving robust statistics for Earth analogs in the primary Cygnus field, but its transit observations have shown statistically that a substantial fraction of the stars in our galaxy have planetary systems and that small rocky planets are ubiquitous \citep{Dressing2015,Buchhave2014,Fressin2013,Howard2012}. Upcoming space missions including the Transiting Exoplanet Survey Satellite \citep[TESS;][]{Ricker2014}, the CHaracterizing ExOPlanet Satellite \citep[CHEOPS;][]{Fortier2014}, and PLAnetary Transits and Oscillation of stars \citep[PLATO;][]{Rauer2014} will detect transiting planets with small radii in short period orbits around bright nearby stars, which will be well-suited for RV follow-up.

There have been several improvements to RV precision over the past two decades. \citet{Butler1996} ushered in an era of 3-\ms\ precision, and the HARPS spectrograph \citep{Mayor2003,Pepe2002} reached even greater RV precision with a vacuum-enclosed, thermally stabilized instrument. There has been significant progress on many of the challenges associated with instrumental stability \citep{Podgorski2014}, and the current state-of-the-art RV precision is now about 1 \ms\ \citep{Fischer2016}. However, this is a factor of ten larger than the RV amplitude for a single Earth-mass planet orbiting a 1 M$_\odot$ star at 1 AU in a circular orbit. Next-generation stabilized spectrographs with ultra-high spectral resolution, laser frequency comb calibration, and improved CCD detectors will aim to reach an instrumental measurement precision of about 10 \cms\ \citep{Jurgenson2016,Halverson2016,Pepe2014}.

These instruments will only succeed if we are able to distinguish stellar photospheric velocities (often collectively called ``stellar jitter'') from orbital velocities. Photospheric velocities manifest themselves as time-correlated red-noise superimposed on Keplerian signals caused by planets. The amplitude of these velocities range from 1 \ms\ for quiet stars to several hundreds of \ms\ for the most active stars. Currently, astronomers try to decorrelate the photospheric contributions to the radial velocity using diagnostic information such as the line bisector span \citep[``BIS SPAN''; as defined in][]{Queloz2001} or FWHM of the cross correlation function, or emission in spectral lines that form in the lower chromosphere such as Ca II H\&K or H-alpha line-core emission. This approach works reasonably well for quiet stars with planets whose orbital velocity amplitudes are greater than 1 \ms, but it has not been successful at disentangling the relative contributions from smaller amplitude signals \citep{Dumusque2017}.

One possible path forward is to use the $\sim$10$^5$ pixels that compose a spectrum to characterize the apparent RV shift due to photospheric velocities instead of trying to decorrelate a post-processed radial velocity measurement based on a global spectral shift. Such a technique could take advantage of the varying sensitivity of specific spectral lines to photospheric effects, as well as subtle line-shape distortions that can not be recognized from a single line. In this work, we apply principal component analysis to simulated spectra to demonstrate under controlled conditions that the spectral signatures of planets and stellar activity features are unique, and that they are imprinted differently in stellar spectra. Our results suggest that there is information embedded in spectra that has gone unutilized by the radial velocity community, and that future statistical techniques could leverage this information to obtain far more precise and accurate RV measurements.

In Sections \ref{sec:jitter} and \ref{sec:sim_spec} we provide an overview of photospheric velocities and present our model to produce simulated active spectra. We then introduce principal component analysis (PCA) in Section \ref{sec:pca} and explore the effects of varying the signal-to-noise (S/N) and instrumental resolution on the PCA results. Finally, we discuss the implications of these results in Section \ref{sec:discussion}.

\section{Photospheric velocities}
\label{sec:jitter}

Stellar RV ``jitter'' is caused by a variety of physical processes. Cool stars have convective envelopes that support acoustic modes with meter-per-second velocity variations on timescales of several minutes \citep{Kjeldsen1995}. Granulation in the photosphere is a manifestation of thousands of rising warm gas cells surrounded by a network of descending cool gas \citep{DelMoro2004}. Granulation flow velocities are \kms, leading to a net blueshift of hundreds of \ms\ in full-disk observations of Sun-like stars \citep{Meunier2017,Gray2009}. The granulation blueshift depends on stellar properties and for a given star varies by meters per second as photospheric magnetic fields evolve over timescales shorter than a few days \citep{Dumusque2011,Lefebvre2008}.

\begin{figure}[h!]
\begin{center}
\includegraphics[width=1.0\columnwidth]{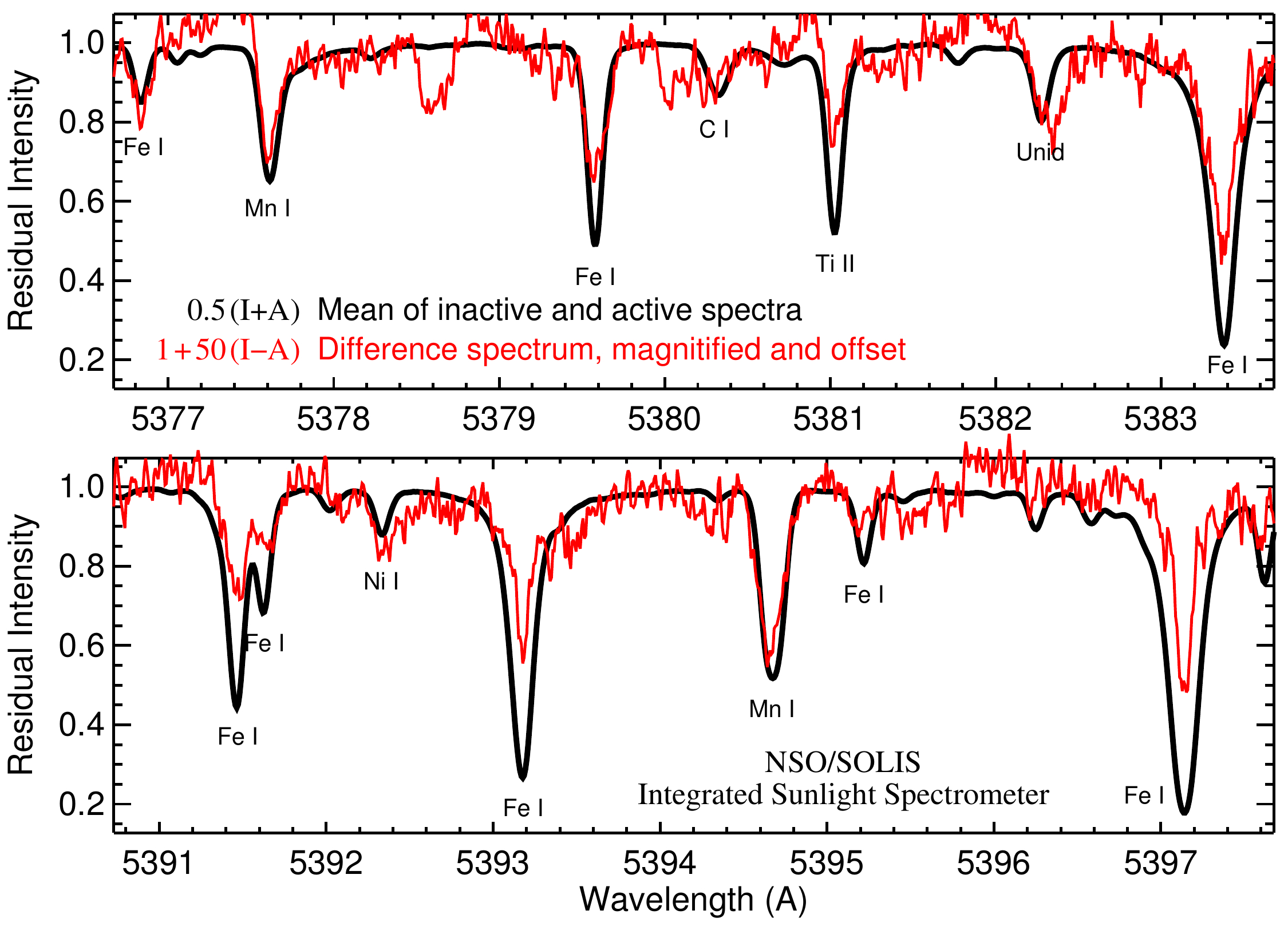}
\caption{{Comparison of the average solar spectrum (black) and the scaled difference between the active and inactive solar spectrum (red) in two nearby bands. Most of the absorption lines seen here are partially filled in by the activity, but there are subtle differences in the way each line responds. Figure courtesy of J. Valenti.
}}
\label{active_spec_fig}
\end{center}
\end{figure}

Magnetic fields coalesce into flux tubes that are bright when they are small (faculae\footnote{Some previous papers, including \citet{Dumusque2014}, refer to faculae as plages. However, plages are the chromospheric counterparts to photospheric faculae and have a more limited effect, filling in the line cores of specific atomic features such as hydrogren absorption.}) and dark when they are large (spots). These flux tubes form and decay on timescales comparable to the stellar rotation period, which is typically days to weeks. As dark spots and bright faculae evolve and rotate across the visible hemisphere, they alter the weighting of projected velocities. Equatorial rotation velocities of \kms\ give rise to \ms\ perturbations due to evolving spots and faculae \citep{Dumusque2014, Lagrange2010, Saar1997}. In practice, these perturbations are responsible for limiting the RV rms of quiet stars to $\sim$1 or 2 \ms\ (e.g., the Rocky Planet Search, \citealt{Motalebi2015}; or the California Planet Search, \citealt{IsaacsonFischer2010}).

Spots and faculae impact photospheric velocities in two main ways. First, the flux effect is induced by the differential contrast of flux between hot faculae or cool spots and the photosphere; breaking the flux balance between the blueshifted approaching limb and redshifted receding limb creates a time-varying radial velocity signal whose magnitude depends on the $v\sin{i}$ of the star \citep{Saar1997} and the temperature difference between the activity feature and surrounding photosphere, $\Delta T$. The flux effect is thought to be the dominant line-shape perturbation for spots on stars with $v\sin{i} > 8$\ \kms\ 
\citep{Dumusque2014}; the dominant broadening of absorption lines for stars with lower $v\sin{i}$ is due to pressure broadening rather than rotational broadening, and so the flux effect does not strongly perturb the wings of these lines. \citet{Haywood2016} determined that the flux effect and inhibition of the convective blueshift effect contribute about 2.4 \ms\ and 0.4 \ms,\ respectively, to the Sun's RV rms.

Second, the uniform convective blueshift of a star's photosphere may be disrupted by magnetic activity, which will suppress convection, resulting an apparent redshift \citep{Cavallini1985,Dravins1981}. This inhibition of the convective blueshift effect is the dominant RV perturbation of faculae, which have only a weak flux effect \citep{Dumusque2014,Meunier2010a,Meunier2010b}. Solar faculae are observed to have filling factors larger than sunspots by a factor of $\sim$10 \citep{Chapman2001}, and therefore the facular inhibition of the convective blueshift effect tends to be the dominant source of RV jitter for slow rotators over timescales comparable to the stellar rotation period \citep{Haywood2016} and the magnetic cycle period \citep{Meunier2010b}.

Taken together, photospheric velocities will add spurious time-coherent scatter to the center-of-mass Doppler velocities. Fortunately, stellar jitter has some distinct properties that we can exploit:

\begin{itemize}
\item photospheric contributions to jitter (such as from spots and faculae) are often tied to the stellar rotation period, which can be measured or estimated from photometric time series \citep[e.g.,][]{Boisse2011},
\item jitter is not a persistent Keplerian signal---it waxes and wanes on varying timescales \citep[e.g.,][]{Gregory2016}, and
\item the magnetic fields and temperatures associated with photospheric activity have unique spectral signatures. For example, low-excitation-potential lines trace cooler components in the photosphere (e.g., spots), whereas high excitation lines indicate warmer components (e.g., faculae). The cores of very strong lines (Ca II H\&K, Balmer lines) are sensitive to chromospheric heating \citep{Noyes1984}.
\end{itemize}

Studies of stellar jitter thus far have generally tried to decorrelate radial velocities derived with either the iodine technique or cross-correlation, and have neglected the rich information content of spectra. Figure \ref{active_spec_fig} shows an average solar spectrum (black) from the Integrated Solar Spectrograph, and the scaled difference between spectra obtained during active and inactive phases (red). Clearly, the spectral response to activity differs from one spectral line to the next on the Sun. This behavior was also recently observed for $\alpha$ Cen B; \citet{Thompson2017} compared spectra from active and inactive phases of $\alpha$ Cen B and found ``pseudo-emission lines'' that were partially filling in absorption troughs, with morphologies that varied on a line-by-line basis.

Identifying the specific lines that respond strongly to activity and characterizing these responses is beyond the scope of this paper, but the simple fact that there are subtle wavelength-dependent differences between quiet and active solar spectra provides information that can be leveraged to construct an improved method of determining radial velocities.

\section{Simulated Spectra}
\label{sec:sim_spec}

In order to examine the detailed spectral effects of stellar activity in a controlled and interpretable experiment, we use the SOAP 2.0 code to generate a collection of spectra from a star with a spot, a facula, or a pure Doppler shift.

\subsection{SOAP 2.0} 
\label{sec:soap}

We use the Spot Oscillation and Planet code 2.0 \citep[SOAP 2.0;][]{Dumusque2014} to create simulated disk-integrated spectra of a star. SOAP 2.0 is an successor to the original SOAP code \citep{Boisse2012}, which simulated the photometric and RV impacts of starspots (but not of faculae). Although the published SOAP 2.0 code performs its calculations and analyses using a 401-data-point cross correlation function (CCF) for computational efficiency, we have modified it to function with entire $\sim$500,000 data-point spectra.

The SOAP 2.0 code breaks a star's surface into a 300-by-300 grid, placing a quiet solar spectrum \citep{Wallace1998} in each grid box; this spectrum has a resolution of $\sim$1,000,000 and S/N of $\sim$1,000. The \citet{Wallace1998} spectrum is continuum normalized, and the telluric features have been fitted out where possible, although some strong telluric regions have been masked out. For grid boxes designated as spots, SOAP 2.0 inserts a sunspot spectrum \citep{Wallace2005}. No high-resolution atlas of facula spectra exists in the literature, and so grid boxes that contain faculae instead use the spot spectrum whose flux is scaled according to the contrast ratio between the faculae and photosphere.

Once spectra are assigned to a grid box, they are shifted according to their projected rotational velocities. The flux effect and inhibition of the convective blueshift are both applied for active regions. Limb-darkening and limb-brightening (for facula) laws are also applied \citep[see Section 2.3 of][]{Dumusque2014}. We adopt $\Delta T_{\mathrm{spot}} = -663$ K, and $\Delta T_{\mathrm{facula}}$ ranges from 35 to 250 K depending on the facula's limb distance \citep{Meunier2010b}. Finally, SOAP sums the individual spectra from each grid box to obtain an integrated spectrum of the entire disk.

\subsection{Model Spectra Created}
\label{sec:model_spectra}

Nine sets of time-series spectra were produced by SOAP 2.0 in the wavelength range from 3925.87 \AA\ to 6661.54 \AA. The nine sets correspond to nine simple cases:

\begin{itemize}
\item an equatorial spot with either $S = 0.1\%$, $S = 1\%$, or $S = 5\%$
\item an equatorial facula with either $S = 0.1\%$, $S = 1\%$, or $S = 5\%$
\item a planet in a circular orbit with either $K = 1\ \ms$, $K = 10\ \ms$, or $K = 50\ \ms$
\end{itemize}

\noindent where $K$ is the radial velocity semi-amplitude of a planet, and $S$ is the filling factor of an active region given by

\begin{equation}\label{eq:filling_factor}
S = \left(\pi R_{\mathrm{AR}}^{2}/2\pi R_{\star}^{2}\right)\times 100\%,
\end{equation}

\noindent where $R_{\mathrm{AR}}$ is the radius of the active region, and $R_{\star}$ is the stellar radius.

Each set is composed of 25 spectra that are evenly spaced in phase over one solar rotation period of 25.05 d. The inclination of both the stellar rotational axis and the planet's orbit is $90^\circ$. Active regions cross the centerline of the visible hemisphere of the star at a phase of 0.

The sizes of the active regions are chosen to represent a range of realistic sizes. For the active Sun, $S = 0.1\%$ spot coverage is typical, while for a star that would be considered ``active'' for an RV survey, such as $\epsilon$ Eri, spots may cover 1\% of the star \citep{Giguere2016}. Very young, extremely active stars, such as TW Hya, may have spot coverage around $S = 5\%$ \citep{Huelamo2008, Donati2011}. Faculae on stars other than the Sun have not been studied in great detail.

To simulate Doppler shifts arising from a planet, we start with the disk-integrated SOAP 2.0 model of the quiet Sun. A planetary RV curve is computed with a period of 25.05 d in circular orbit. The mass of the planet is selected so that the RV amplitude is similar to the amplitude of the variability from the spots or faculae according to \citet{Dumusque2014}. For each point in the time-series RVs, the shifted wavelengths, $\lambda_s$, are calculated using the relativistic Doppler formula

\begin{equation}\label{eq:shift}
\lambda_s = \lambda_0\ \frac{1+\frac{v}{c}}{\sqrt{1-\frac{v^2}{c^2}}},
\end{equation}

\noindent where $\lambda_0$ is the set of original wavelengths, $v$ is the RV, and $c$ is the speed of light in a vacuum \citep{Einstein1905}. In order to apply principal component analysis to this data set, it is necessary to resample the shifted spectrum from $\lambda_s$ back to $\lambda_o$ (see Section \ref{sec:ideal_pca}) with cubic spline interpolation.

These SOAP 2.0 integrated spectra (with no added noise and with full resolution) are labeled as our ``ideal" spectra; they are used as the starting point for creating more realistic simulated spectra with a range of spectral resolutions and S/N. For every S/N and resolution combination we choose, we create fifty sets of spectra with independent realizations of noise.

Resolution $R$ is obtained by convolving with a Gaussian whose FWHM is given by

\begin{equation}\label{eq:fwhm}
\mathrm{FWHM}(\lambda) = \lambda/R.
\end{equation}

\noindent The average S/N per resolution element is

\begin{equation}\label{eq:snr}
S/N = (S/N)_{\mathrm{px}}\times\sqrt{s},
\end{equation}

\noindent and we adopt $s = 3$ for the sampling of the line spread function.

Our realistic simulated spectra do not include other effects such as the S/N loss from the blaze function, or lower throughput of blue wavelengths (e.g., EXPRES, \citealt{Jurgenson2016}; or HARPS, \citealt{Mayor2003}). We also ignore the effect of time-varying telluric contamination.

\section{Principal Component Analysis}
\label{sec:pca}

Principal Component Analysis (PCA; also called the Karhunen-Lo\`{e}ve transform in certain applications) is a standard statistical technique with a variety of applications \citep{Pearson1901}. It can be used to reconstruct data based on a small number of principal components to denoise spectra \citep{MartinezGonzalez2008} or for processing high-constrast images \citep{Soummer2012}. PCA has also been used to measure line-shape perturbations in spectral lines in order to estimate the average magnetic field strength of a star \citep{Lehmann2015}, and to explore the impact of stellar activity on the CCF \citep[see Section 4.2.2 of][]{Fischer2016}.

Given an $n \times p$ data matrix $Y$, PCA is a process of defining a new coordinate system for $Y$ that is made up of orthogonal dimensions representing the directions of decreasing variance in the data. The first dimension of the new coordinate system is labeled as principal component (PC) 1; this is the direction in $p$-dimensional space of greatest variance in the original data. PC 2 is the orthogonal direction that has the second greatest variance, and so forth. This procedure can continue until $p$ PCs have been calculated, but in practice, the majority of the variance in the data matrix is often captured in only $m$ PCs, where $m << p$. When this occurs, PCA can be an effective method for dimension reduction with minimal information loss. 

We perform PCA on a data matrix $Y$, which contains one set of 25 time-series spectra. The $i^{\mathrm{th}}$ row, $j^{\mathrm{th}}$ column element $Y_{ij}$ is the intensity of the $j^{\mathrm{th}}$ wavelength at time $t_i$. $Y$ is column-centered (i.e., column means are set to zero) and is scaled (i.e., column values are divided by their standard deviations). $Y$ is then factorized using singular value decomposition to obtain

\begin{equation}\label{eq:svd}
Y_{n,p} = U_{n,n} \times S_{n,p} \times W_{p,p}^\mathrm{T},
\end{equation}

\noindent where $U$ and $W$ are both orthonormal matrices, and $S$ is a diagonal matrix whose entries are the singular values.

In this factorization, the $k^{\mathrm{th}}$ column of $W$ is the $k^{\mathrm{th}}$ principal component vector. The magnitude of the $j^{\mathrm{th}}$ component of the $k^{\mathrm{th}}$ PC vector indicates the relative amount that the $j^{\mathrm{th}}$ wavelength contributed to the $k^{\mathrm{th}}$ PC direction. In other words, if the $j^{\mathrm{th}}$ PC 1 vector component has a large magnitude, then it indicates that the $j^{\mathrm{th}}$ wavelength is responsible for a large amount of variance in the data.

The ``scores'' for principal component $k$ are the projections of each row of $Y$ onto the PC $k$ direction and are given by $Y_{n,p} \times W_{p,p}$. Therefore, score $k$ represents the relative locations of each spectrum along PC $k$. If a particular spectrum has a score that is far from zero for a given PC, then the spectrum occupies a more extreme position along that PC direction compared to the other spectra in the data matrix.

Since $Y$ is centered and scaled, the $k^{\mathrm{th}}$ PC captures some fraction, $f_{k}$, of the total variance in the data; $f_{k}$ is given by

\begin{equation}\label{eq:frac_of_var}
f_{k} = \frac{S_{k,k}^{2}}{np},
\end{equation}

\noindent where $S_{k,k}$ is an entry in the diagonal matrix $S$. PCA requires that if $k < l$, then $f_{k} \geq f_{l}$, which ensures that the PCs are sorted in order of the amount of variance captured.

Since these are simulated spectra, there is no barycentric correction to apply, and so we do not need to put the spectra into the star's reference frame. With real data, however, it would be necessary to ensure that every spectrum is in the same reference frame so that spectral features are aligned in the data matrix. It is also essential that the spectra in $Y$ be sampled at identical wavelength values because PCA treats each column as an independent variable, and therefore it does not look for any relation between neighboring wavelengths.

\subsection{PCA of Ideal Spectra}
\label{sec:ideal_pca}

\begin{figure}[]
\begin{center}
\includegraphics[width=1.0\columnwidth]{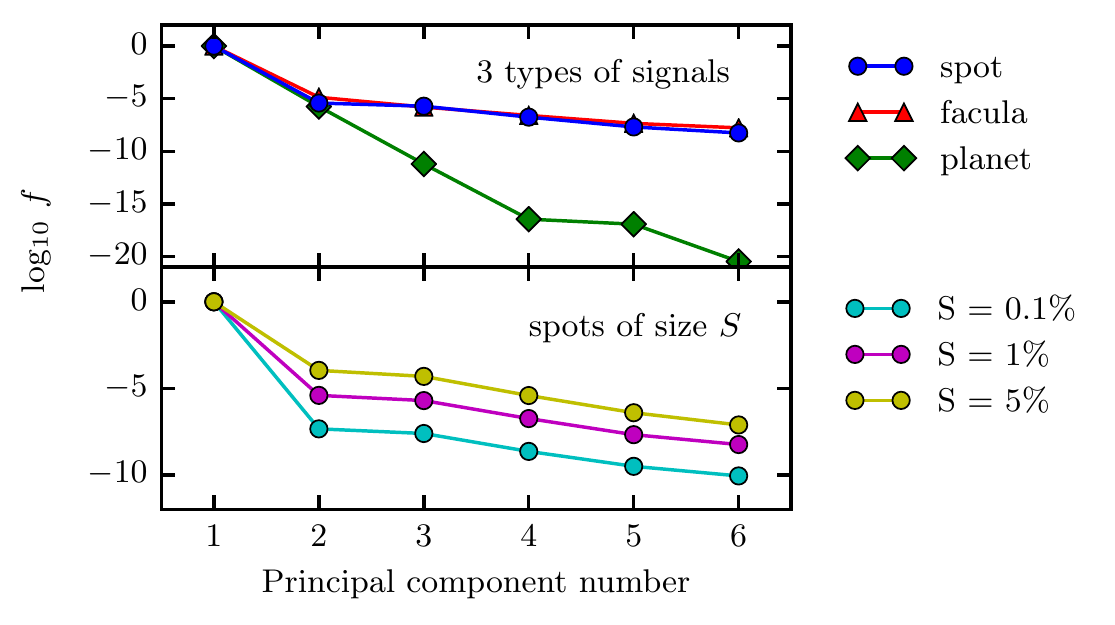}
\caption{{Fraction of variance $f$ captured by the first six principal components. Top: $f$ for an $S = 1\%$ spot, $S = 1\%$ facula, and $K = 10\ \mathrm{m\ s^{-1}}$ Doppler shift. $f$ falls rapidly for the planet, but later PCs are capture more variance for the active regions. Bottom: $f$ for three different sizes of injected spot signals. Larger spots have more variance captured in later PCs. Similar results are found for faculae of varying size.
}}
\label{PCA_fracs_fig}
\end{center}
\end{figure}

We use PCA to examine the ideal (i.e., no noise added and with full resolution) SOAP 2.0 spectra sets. Figure \ref{PCA_fracs_fig} shows the fraction of variance $f$ captured by each PC for a number of cases. We find that PC 1 captures more than 99.99\% of the variance in every set. For the spots and faculae, subsequent PCs do offer some information, while the higher PCs for planets have far smaller $f$ values. In our simulated data, there is real information contained beyond PC 1 in the case of spots and faculae, with higher principal components capturing more variance for the larger activity features.

\begin{figure*}[]
\begin{center}
\includegraphics[width=\linewidth]{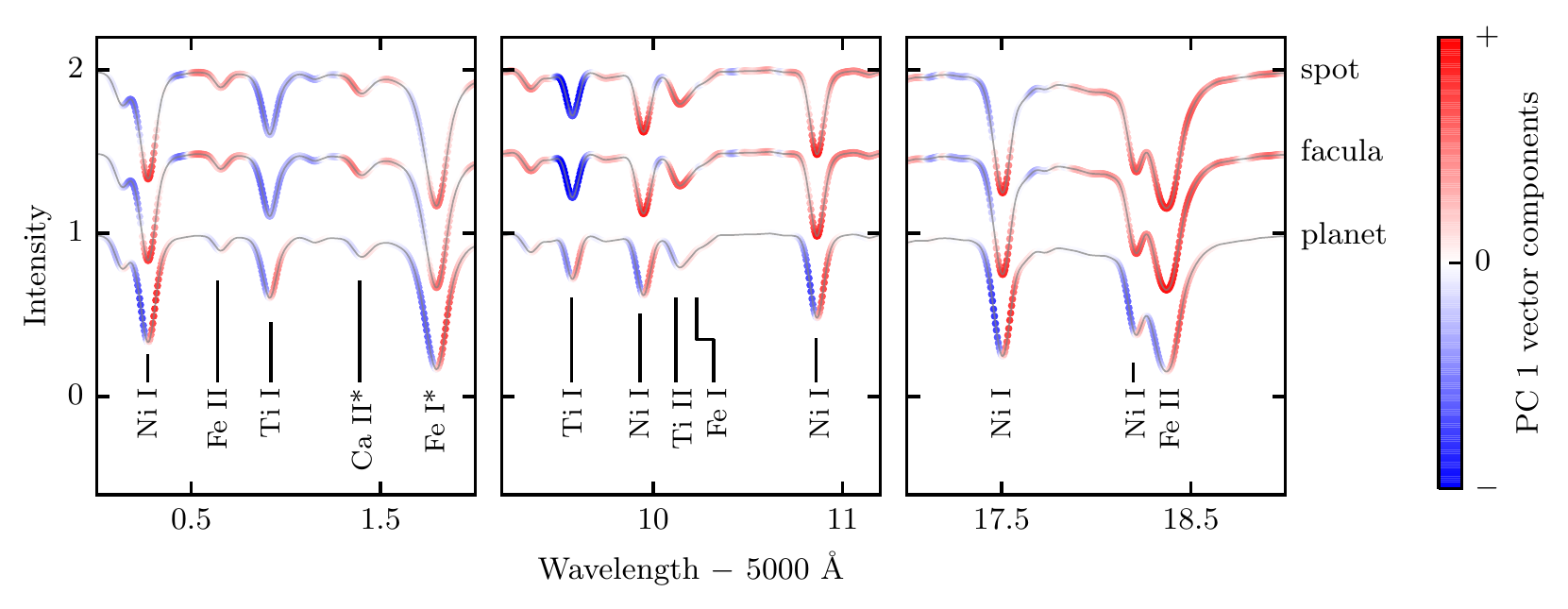}
\caption{{Normalized spectra and associated PC 1 vector components. A thin gray line traces one spectrum for the $S= 1\%$ spot, $S= 1\%$ facula (offset by $+0.5$), and $K = 10\ \mathrm{m\ s^{-1}}$ planet (offset by $+1$) sets at maximum S/N and resolution. Overplotted colored points represent the values of the PC 1 vector components for each set. Blue and red are used for wavelengths that contribute to opposite directions of variation along the PC 1 axis, with white representing wavelengths that contribute minimal variance along that axis. Some spectral lines are labeled. An asterisk denotes a line blend.
}}
\label{PC_vectors_fig}
\end{center}
\end{figure*}

Figure \ref{PC_vectors_fig} examines the structure of PC 1 vector components for these same three sets. The magnitudes of the planet's PC vector components are greatest where the slopes of the spectral lines are greatest, since these are the wavelengths that experience the greatest variation when the spectra are redshifted and blueshifted. As a result, the structure of the PC 1 vector components is qualitatively identical for every single line in the Doppler-shifted spectra, unlike for the active region spectra, whose vector components differ from line to line. This demonstrates that the spectral variability is manifested very differently for spectra with active regions than for pure Doppler shifts.

There are several examples of lines that vary greatly in the spot and facula sets. Both of the Ti I lines in Figure \ref{PC_vectors_fig} show high variance in a particular PC 1 direction (shown as blue). The Ni I line near 5011 \AA\ responds in the opposite PC 1 direction (shown as red). This window was chosen arbitrarily, and there are numerous examples of strongly responsive wavelengths across the entire spectrum.

\begin{figure}[]
\begin{center}
\includegraphics[width=1\columnwidth]{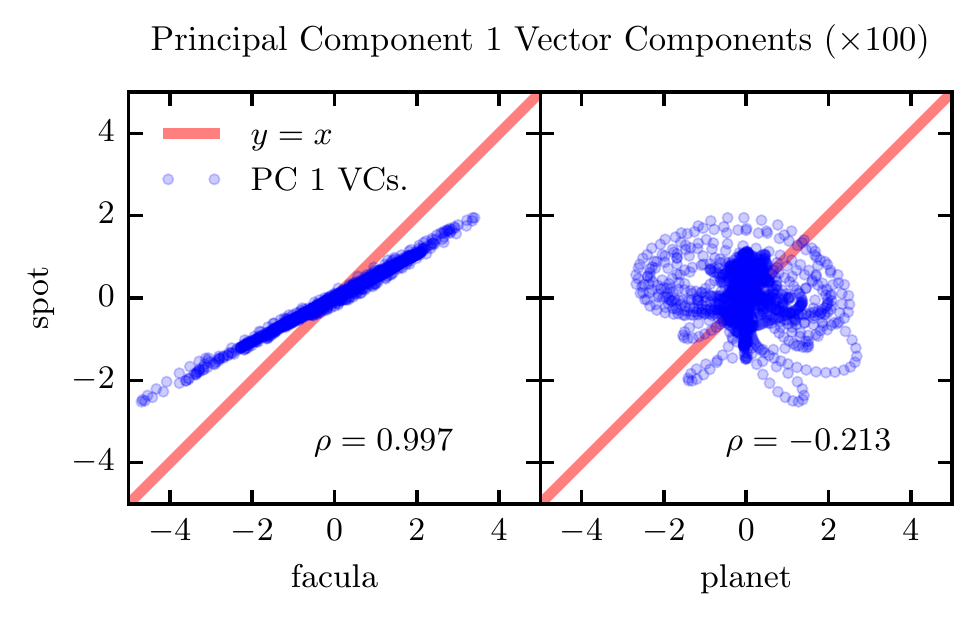}
\caption{{Correlations between the PC 1 vector components of SOAP 2.0 spectra sets from 5000 to 5010 \AA. PC vector components plotted here correspond to the same spectra shown in Figure \ref{PC_vectors_fig}. The high Pearson product-moment correlation coefficient, $\rho$, shows that the PC 1 vector components for the $S=1\%$ spot and $S=1\%$ facula sets exhibit a tight, linear relationship (left). By contrast, the PC 1 vector components for the spot and planet sets show no significant correlation (right).
}}
\label{PC_vectors_corr}
\end{center}
\end{figure}

The PC 1 vector components for the spot and facula sets are nearly indistinguishable in Figure \ref{PC_vectors_fig}. Figure \ref{PC_vectors_corr} verifies that these vector components are extremely well-correlated with one another, but not with the PC 1 vector components for the planet. This implies that the variability in the spot and facula sets is extremely similar (modulo scaling), while the spectra of the active regions and the planet vary differently. A likely explanation for this correlation is that SOAP 2.0 uses the sunspot spectrum as a starting point when producing both spots and faculae; it is possible that the spectral alterations applied by SOAP 2.0 for the facula are small compared to the intrinsic line-by-line variability between a spectra of a spot and the quiet photosphere. 

\subsection{PCA of Realistic Simulated Spectra}
\label{sec:real_pca}

We use our realistic simulated spectra to explore the relation between S/N, resolution, and the information content of active region spectra and pure Doppler-shifted spectra. For the $j^{\mathrm{th}}$ realization of noise, score $i$ is calculated for a realistic spectra set; this is labeled $Z_{ij}$. The structure of $Z_{ij}$ as a function of time is compared to the structure of score $i$ for the corresponding ideal spectra set, $Z^{\mathrm{0}}_i$.

\begin{figure}[]
\begin{center}
\includegraphics[width=1\columnwidth]{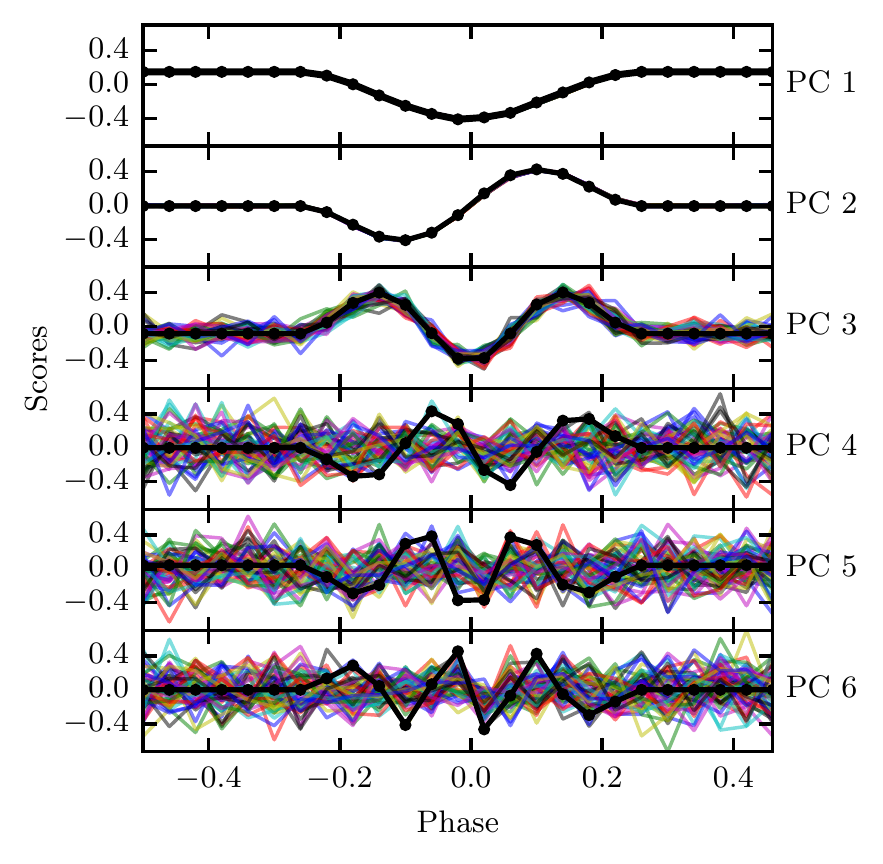}
\caption{{Scores 1 through 6 for fifty realizations of noise for the 1\% spot series. For this example, the resolution was $R = 150,000$, and the $\mathrm{S/N} = 800$. Colored lines represent the scores of individual noise realizations, while the black lines indicate the scores of the ideal spectra with full S/N and resolution. Visual inspection suggests that scores 1 through 3 show a high degree of correlation between the realistic spectra and the ideal spectra (cf. Figure \ref{rho_p_histo_fig}).
}}
\label{scores_fig}
\end{center}
\end{figure}

As noise is added and as the resolution is reduced, scores corresponding to earlier PCs maintain their structure, but the scores for higher PCs eventually become noise dominated. This trend is demonstrated in Figure \ref{scores_fig}, which shows scores 1 through 6 for the $S = 1\%$ spot at $R = 150,000$ and $\mathrm{S/N} = 800$. It is clear in this example that for scores 1 through 3 there is close agreement between $Z^{\mathrm{0}}_i$ and the scores of the fifty noise realizations for the realistic spectra. For scores 4 through 6, there is no such agreement.

We quantify the closeness of this agreement for the score $i$ and noise realization $j$ by calculating the Pearson product-moment correlation coefficient, $\rho_{ij}$, of $Z_{ij}$ and $Z_{i}^{0}$. Since the sign of the PC directions and scores are arbitrary in PCA, we consider only the absolute value of each $\rho_{ij}$ when we assess the strength of the correlation between $Z_{ij}$ and $Z^{\mathrm{0}}_i$. We also compute the $p$-value for each correlation in order to test the null hypothesis that the correlation between $Z_{ij}$ and $Z^{\mathrm{0}}_i$ is zero against the alternative that it is not zero (i.e., a two-sided alternative). \footnote{The Fisher Transformation was used on the correlation coefficients as the test statistic; when $Z_{ij}$ and $Z^{0}_i$ are close to Gaussian, the Fisher Z Transformed correlation's sampling distribution is approximately Gaussian.}

\begin{figure*}[]
\begin{center}
\includegraphics[width=0.9\linewidth]{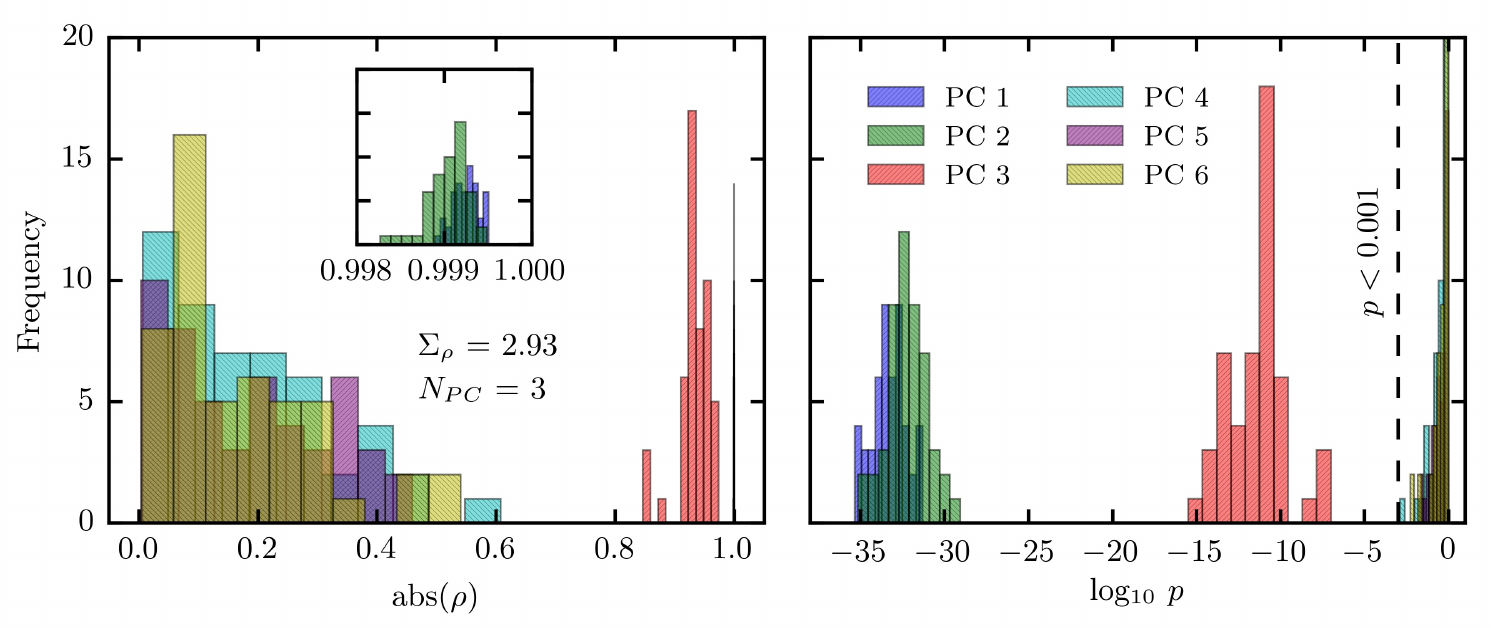}
\caption{{Left: Histograms of the correlations between the scores of the ideal spectra set for a 1\% spot and scores for fifty realizations of realistic spectra sets with $\mathrm{S/N} = 800$ and $R = 150,000$. The inset magnifies the region near unity for score 1 and score 2. For this collection, $\Sigma_{\rho} = 2.93$: score 1 and score 2 each contribute a value of $\sim$1, and score 3 contributes $\sim$0.9. There are therefore three significant PCs. Right: Histograms of $p$-values for the same set. The $p$-values for scores 4 through 6 are all greater than the cut-off of 0.001, and so they do not contribute at all to $\Sigma_{\rho}$.
}}
\label{rho_p_histo_fig}
\end{center}
\end{figure*}

Figure \ref{rho_p_histo_fig} shows the distributions of $|\rho|$ and the $p$-values for the case of the $S = 1\%$ spot at $R = 150,000$ and $\mathrm{S/N} = 800$. As in Figure \ref{scores_fig}, it is evident that the score 1 and score 2 are extremely well-correlated between ideal and realistic sets, with subsequent scores showing less correlation. Scores 4 through 6 have correlation distributions that are peaked near zero, and therefore are unlikely to contain real structure. This is captured as well by the $p$-value distributions for scores 4 through 6, which are significantly greater than $p = 0.001$.

We define a quantity $\Sigma_{\rho}$ to represent the number of PCs whose scores can be recovered with confidence for a particular S/N and resolution. We allow $\rho_{ij}$ values to contribute towards $\Sigma_{\rho}$ only if the corresponding $p$-values, $p_{ij}$, are less than 0.001.\footnote{Since multiple hypothesis tests are carried out, the $p$-value cut-off of 0.001 is not the true significance level. There are various ways to account for multiple testing. For each PC, fifty tests are run. A simple, though conservative, adjustment is the Bonferroni correction, which would give a family-wise error rate of $0.001\times 50 = 0.05$.} We define a function $g$ to enforce this condition: 

\begin{equation}\label{f_function}
g(p) \equiv \left\{
     \begin{array}{lr}
       1 & : p < 0.001\\
       0 & : p \geq 0.001
     \end{array},
   \right.
\end{equation}

\noindent where $p$ is a $p$-value. We can then define $\Sigma_{\rho}$ as

\begin{equation}\label{sigma_rho}
\Sigma_{\rho} \equiv \sum\limits_{i=1}^{10} \sum\limits_{j=1}^{50} g(p_{ij}) \frac{\rho_{ij}}{50},
\end{equation}

\noindent where $i$ is the index over the 10 PCs that were computed for each set, and $j$ is the index over the fifty realizations of noise for each set.

Finally, we define an integer quantity, $N_{PC}$, which is equal to $\Sigma_{\rho}$ rounded to the nearest integer (with 0.5 rounding to 1). $N_{PC}$ will serve as a metric for comparing the relative information content of a set of spectra with varying S/N and resolution.

\begin{figure*}[]
\begin{center}
\includegraphics[scale=0.9]{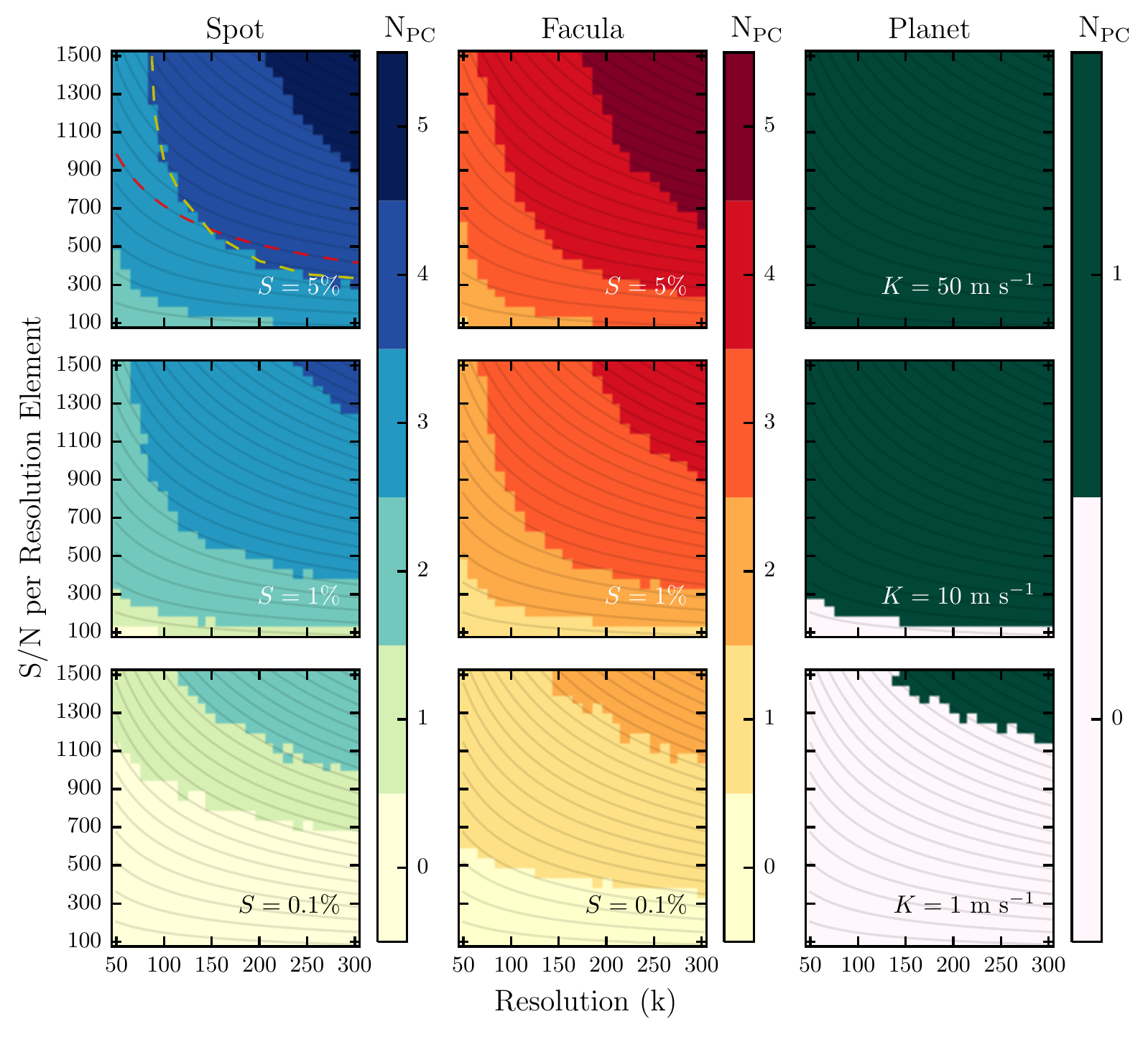}
\caption{{$N_{PC}$ for sets of spectra that contain a spot, facula, or Doppler shift of varying sizes. Within each subplot, the instrumental resolution and S/N per resolution element are varied. Regions of parameter space that have high $N_{PC}$ values contain more information in their spectra than those with lower $N_{PC}$ values within the same family of features (e.g., $S = 1\%$ spots). Gray lines are lines of equal photon flux; as light is dispersed at higher resolution the S/N falls correspondingly.
}}
\label{N_PC_fig}
\end{center}
\end{figure*}

Figure \ref{N_PC_fig} shows how $N_{PC}$ varies as a function of S/N, instrument resolution, and the size of the activity feature or Keplerian RV amplitude. The lines of equal photon flux in Figure \ref{N_PC_fig} indicate the expected relation between S/N and resolution for a given amount of flux and a fixed sampling:

\begin{equation}\label{eq:snr_R}
\mathrm{S/N} \propto \frac{1}{\sqrt{R}}.
\end{equation}

\noindent This relation holds in the photon-limited observational regime considered in this work. For example, HIRES ($R$ = 55,000) and HARPS ($R$ = 115,000) each obtain typical S/N of a few hundred (see \citealt{Fischer2016} for the resolution and typical S/N of many other current RV instruments).

Comparing the lines of equal photon flux to the $N_{PC}$ breakpoints reveals that high resolution is important for identifying photospheric signals, providing larger $N_{PC}$ values even after accounting for the concomitant S/N decrease. An example is shown in the $S = 5\%$ spot subplot: a particular line of equal photon flux (red dashed line) crosses the breakpoint between $N_{PC} = 3$ and $N_{PC} = 4$ (yellow dashed line) near a resolution of 150,000. In general we see that breakpoint crossings occur at higher resolutions for lower S/N values.

The three Doppler-shift cases shown in Figure \ref{N_PC_fig} look completely different from their active region counterparts, even though the effective RV semi-amplitudes of the sets are similar. For even the largest pure Keplerian signals examined, there is at maximum only one significant PC. Noise becomes dominant for the $K = 1$ \ms\ signal over much of parameter space, yielding $N_{PC} = 0$. 

\section{Discussion}
\label{sec:discussion}

Our simulations show that PCA reveals variability in time-series spectra that is correlated with the presence of spots, faculae, or planets. This work examines the isolated effects of these phenomena as a first step in learning how to disentangle the more realistic case of combined spots, faculae, and planetary signals. In this section we review our results and discuss them in the context of moving towards this goal.

\subsection{Spectral-Line Dependence of Activity}
\label{sec:spec_line_dep}

We find that the directions and magnitudes of variance (i.e., the principal component vector components) in time-series spectra of a spot or facula are significantly different than those corresponding to spectra containing a Doppler shift. The PC 1 vector components for activity features show structure that varies from one spectral line to another; this wavelength dependence is distinct from the broad wavelength dependence related to the contrast ratio between active regions and photosphere \citep[c.f.,][]{Reiners2010}. We interpret this line-by-line difference as arising from the varying sensitivity of specific atomic transitions to temperature variations, or to the depth of formation in the photosphere. Ti I, for instance, is a temperature-sensitive transition; we posit this sensitivity is the reason for the unique structure of Ti I's PC 1 vector components in Figure \ref{PC_vectors_fig}. This type of line-by-line information has not yet been fully exploited by current RV techniques, and our results show that there is a wealth of information hidden within the thousands of individual spectral lines.

Our observations of line-by-line spectral variability are similar to those of T. Carroll, whose work is described in Section 4.2.2 of \citet{Fischer2016}. Carroll used PCA to analyze the CCFs of HARPS spectra of the slow rotator HD 41248. Carroll found that PC 1 contained nearly the entire Doppler signal, and that the Doppler signal's amplitude varied by $\sim$150 \ms\ comparing CCFs derived from high- or low-excitation-potential lines.

The difference between the PC 1 vector components of Doppler shifts and stellar activity features, illustrated in Figures \ref{PC_vectors_fig} and \ref{PC_vectors_corr}, demonstrates that spectra contain the information needed to distinguish between these phenomena if the spectral resolution and S/N are high enough. Figure \ref{eq:frac_of_var} shows that the variance in these spectra can be described with a small number of PCs, meaning that information that can be combined across all of these wavelengths to reveal new, vector-based stellar activity indicators that may well be more informative than traditional scalar indicators (e.g., CCF FWHM or BIS SPAN).

Our results imply that current methods to decorrelate RVs have room for improvement, and that statistical techniques leveraging the pixel-by-pixel variability in time-series spectra offer a promising path forward. The current state-of-the-art RV technique involves deriving raw RV measurements from the center of the CCF, and then correcting these RVs based on activity indicators \citep[using, e.g., Gaussian Processes;][]{Rajpaul2015}. We show that absorption lines respond to activity in a non-uniform way, and therefore, averaging over thousands of lines that have each been perturbed by stellar activity will necessarily washout information. Activity indicators based on the CCF's shape are also based on the average perturbation of absorption lines, and so these too are diluted by the line-by-line variability of stellar activity. Furthermore, non-CCF-based activity indicators, such as Ca II H\&K or H-alpha line-core emission, are created in the chromosphere, and are therefore only imperfectly correlated with the instantaneous photospheric velocity fields, which are the true cause of spurious RV signals. Using PCA and controlled simulated spectra, we have shown that it is possible to empirically quantify the variability in each of the $\sim$10$^5$ pixels composing a spectrum, which is a direct probe of the spectral manifestations of photospheric activity.

\subsection{The Value of High Resolution}
\label{sec:value_high_R}

Comparing the structures of the principal component scores for spectra with lower resolution and S/N to the ideal spectra reveals that higher resolution is better able to retrieve information content from spectra that have been affected by stellar activity. Figure \ref{N_PC_fig} demonstrates in the $S = 1\%$ spot case, for instance, there are certain regimes in which increasing the resolution will permit more significant principal components to be recovered, and therefore greater information content, despite suffering from the accompanying S/N loss. Higher resolution requires longer exposure times to reach a given S/N. Our simulations also show that additional principal components can also be seen with higher S/N. In an era where stellar magnetic activity is the main obstacle in detecting low-mass planets, it will be very beneficial to consider the trade-off between S/N and resolution at the design phase for an instrument. The exposure time scales linearly with increasing resolution, but, of course, exposure time scales as the square of the S/N.

The advantages of higher resolution are at odds with previous studies that report diminishing returns in RV precision beyond $R \sim$ 100,000 \citep[e.g.,][]{Bouchy2001}. Because the RV precision is proportional to the slope of the spectral lines, the precision does not improve significantly once the spectral lines are fully resolved at a resolution of about 80,000. However, these simulations ignore the impact of photospheric velocities. \citet{Dumusque2014} demonstrated that active regions on slowly rotating stars produce line profile variations because of convective blueshift inhibition. High resolution better samples the line profile, and therefore it provides information that can be used to better characterize stellar activity. Our simulations show that this information is still imprinted in the spectrum and that, with high resolution, it is possible to distinguish these line variations from Keplerian Doppler shifts. 

With a new generation of high-resolution spectrographs imminent, this result is encouraging for future studies of young and active stars, whose planetary populations have so far been exceptionally difficult to probe with the radial velocity technique because of stellar jitter on the order of hundreds of \ms. Given the significant and distinctive signatures that large photospheric features have displayed in our simulated spectra, it seems plausible that RV jitter could be reduced around these active stars with next-generation high-resolution spectrographs and newly developed statistical techniques.

The ultimate goal, of course, it is to disentangle the simultaneous effects of sub-meter-per-second Doppler shifts and of small additional spot and facula perturbations. Figure \ref{N_PC_fig} shows that the $S = 0.1\%$ spot and facula cases and the $K = 1$ \ms\ planet case all have $N_{PC} < 1$ over much of the parameter space occupied by many current and future planet-searching spectrographs. This result need not be concerning, however, because this only implies that the variance due to noise is greater than the variance due to the injected signal in an individual spectrum. In reality, $\sim$1-meter-per-second planets are detectable because analysis methods are designed to search for Keplerian shifts and dozens to hundreds of observations are used to recognize the periodic signal. We are optimistic that new statistical techniques may prove similarly successful for activity features once the full information content of the spectrum is utilized.

\section{Conclusion}

This paper presents our application of principal component analysis (PCA) to examine the spectral signatures of spots, faculae, and pure Doppler shifts in simulated spectra produced with the SOAP 2.0 code. Our motivation is to move towards the development of a new method of computing Keplerian radial velocities that utilizes the rich information content of the $\sim$10$^5$ pixels constituting a spectrum to fit simultaneously for both Doppler shifts due to planets and spectral-line perturbations that are astrophysical in origin.

We applied PCA to disk-integrated time-series spectra of spots and faculae to reveal that their spectral signatures are distinct from those of planets. While a set of Doppler-shifted spectra shows qualitatively similar variability for every line, each absorption line in the active spectra is affected differently; this could lead to the identification of new indicators that directly probe photospheric activity. In our simulations, we found that the information required to distinguish photospheric and planetary signals is contained within the stellar spectrum, and that it should be possible to exploit this information with high-quality data and an appropriate statistical framework.

When we applied PCA to spectra with realistic instrumental resolution and noise, we found that a number of the principal components were still nearly identical to those of the ideal spectra. Through this simulation we also found that stellar activity features are described by multiple significant principal components (especially larger features), while Doppler-shifted spectra are described by only one significant principal component. According to our simulations, extremely high resolution, even in excess of $R$ $\sim$ 150,000, gives a comparative advantage over high S/N when attempting to maximize the information content in observations that contain photospheric activity. The subtle effects of photospheric activity are contained in the profiles of absorption lines, and high resolution gives additional information about higher-order spectral variability that may be essential as we move towards the more complex case of combined stellar activity and planetary signals.

To fully take advantage of upcoming survey missions like TESS, it is essential that we overcome stellar noise so that small, nearby planets can be characterized. Our work suggests that statistical techniques operating on a pixel-by-pixel basis on high-quality data from next-generation spectrographs will offer a promising path forward towards measuring and correcting for photospheric velocities.

\section{Acknowledgements}

We are grateful to our referees for providing thoughtful and constructive feedback on this paper, which has greatly improved its presentation. We thank Jeff Valenti for providing Figure \ref{active_spec_fig}. We also acknowledge helpful conversations with Lars Buchhave. The authors acknowledge support from the NSF grant AST1616086. A.B.D. acknowledges support through the NSF Graduate Research Fellowship grant DGE1122492. X.D. acknowledges support from the Society in Science--The Branco Weiss Fellowship, and the Swiss National Science Foundation for funding through the National Centre for Competence in Research ``PlanetS.'' E.B.F. acknowledges support by the Pennsylvania State Office of Science Engagement and the Center for Exoplanets and Habitable Worlds, which is supported by the Pennsylvania State University, the Eberly College of Science, and the Pennsylvania Space Grant Consortium. E.B.F. also acknowledges support from NASA Exoplanets Research Program \#NNX15AE21G. This work was partially supported by the NSF grant DMS-1127914 to the Statistical and Applied Mathematical Sciences Institute (SAMSI), and E.B.F. and D.F. acknowledge supporting collaborations within NASA's Nexus for Exoplanet System Science (NExSS).

\bibliographystyle{aasjournal.bst}
\bibliography{PCA_detectability}

\end{document}